\documentclass[twocolumn,showpacs,%
  nofootinbib,aps,superscriptaddress,%
  eqsecnum,prd,notitlepage,showkeys,10pt]{revtex4-1}

\usepackage{amssymb}
\usepackage{amsmath}
\usepackage{graphicx}
\usepackage{dcolumn}
\usepackage{hyperref}

\begin{document}



\maketitle

{\bf Comment on ``Asking Photons Where They Have Been.''}

In a recent letter, A. Danan et al. \cite{Danan} devised an elegant experiment investigating the past of photons inside two Mach-Zehnder interferometers, one inside the other---yet drew the wrong conclusions. Namely, that---based on weak measurements on pre- and post-selected states---some photons have been inside the inner interferometer but they never entered and never left. And, consequently, a ``common sense'' approach describing the past of a photon in terms of a trajectory, or a set of trajectories \cite{Salih}, \cite{Hosten}, should be abandoned.

\begin{figure}
\centering
\includegraphics[width=0.5\textwidth]{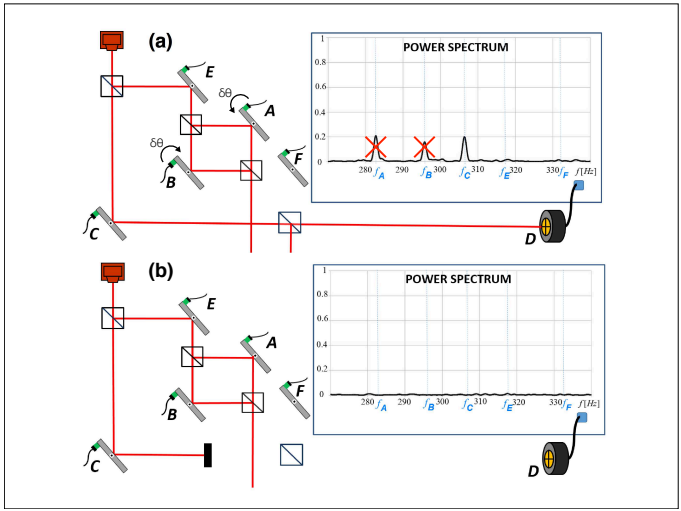}
\caption{\label{fig: figure}(a) If mirrors A and B are made to vibrate at exactly the same frequency such that whenever A is rotated by a small angle $\delta \theta$, B is rotated by $-\delta \theta$, then complete destructive interference at F is preserved. The peaks corresponding to the vibrational frequency of mirrors A and B should now disappear. The photons have not been inside. (b) For the case when the lower path is blocked, again standard quantum mechanics fully and simply explains the absence of any peaks, which the TSVF does not do.}
\end{figure}

But if weak measurements are performed such that \emph{zero} photons leave the inner interferometer---that is complete destructive interference is not disturbed---these measurements do not find the photons to have been inside it. We show, using two key setups discussed by the authors, FIG. \ref{fig: figure}, that standard quantum mechanics not only explains the physics better than the two-state vector formulation (TSVF) the authors advocate, but that it can tell a very different story.

Let's first consider the  setup of FIG. \ref{fig: figure} (a) based on which the authors draw their main conclusions. By vibrating the mirrors, the authors were able to cause light reaching the quad-cell photo-detector D to acquire vertical shifts, each significantly smaller than the beam width---essentially performing weak measurements on the pre-selected state $\frac{1}{\sqrt{3}}(\left| \text{A} \right\rangle+i\left| \text{B} \right\rangle+\left| \text{C} \right\rangle)$, and the post-selected state $\frac{1}{\sqrt{3}}(\left| \text{A} \right\rangle-i\left| \text{B} \right\rangle+\left| \text{C} \right\rangle)$. The presence of peaks corresponding to the vibrational frequencies of mirrors A and B indicates that some photons have been near these mirrors.

\begin{figure}
\centering
\includegraphics[width=0.5\textwidth]{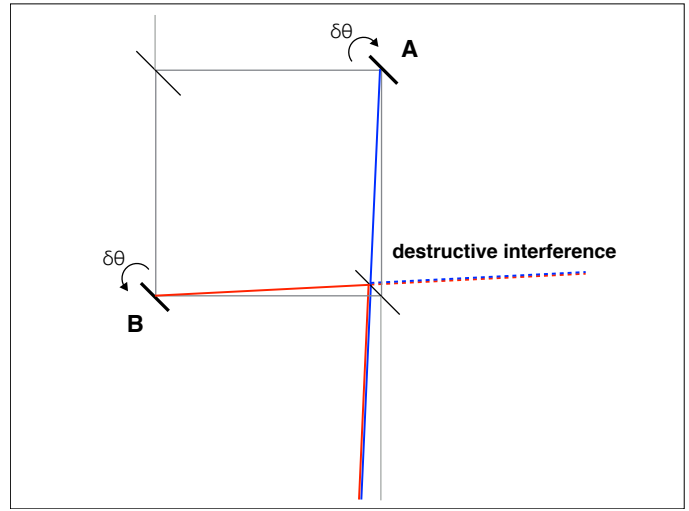}
\caption{\label{fig: sketch}Rotating mirror A by a small angle $\delta \theta$, and mirror B by $-\delta \theta$, does not disturb complete destructive interference as the signals from A and B reach the bottom-right beam-splitter at exactly the same position, and the reflected part of the signal from A and the transmitted part of the signal from B (two dotted lines) overlap exactly. The distances from A and B to the bottom-right beam-splitter are assumed to be equal. Any required phase shift may be implemented as a delay. Note that in this drawing $\delta \theta$ is exaggerated for clarity.}
\end{figure}

Now imagine that both mirrors A and B are made to vibrate at exactly the same frequency, such that whenever A is rotated by a small angle $\delta \theta$, B is rotated by $-\delta \theta$, i.e. in the opposite direction. By simple geometry, and provided that the distances from A and B to the beam-splitter on the way to F are equal, it is easy to see that complete destructive interference at F is not disturbed  \cite{Contrary}, FIG. \ref{fig: sketch}.

What happens to the peaks corresponding to mirrors A and B? The weak values of the projection operators at A and B are +1 and -1 respectively. The minus sign for mirror B indicates an average vertical shift for the photons arriving at the detector that is in the opposite direction of that associated with A---the two thus cancelling each other out. The peaks corresponding to the vibrational frequency of mirrors A and B now disappear.

The story told by the TSVF is the following. Since the forward-evolving state (from the source) and the backward-evolving state (from the detector) are both non-zero at mirror A and at mirror B, the photons have been there. But because the physical effects at the detector of the two weak measurements A and B are opposite, we observe no net effect, raising the question: were the photons shifted at the detector in one direction due to A and simultaneously shifted in the opposite direction due to B?---whose answer has to be no since a joint weak measurement at A and B, to see if the photons were on paths A and B simultaneously, gives the result zero. One can see this from the fact that $\left| \text{A} \right\rangle \left\langle \text{A} \right| \left| \text{B} \right\rangle \left\langle \text{B} \right|$ is identically zero, as $\left| \text{A} \right\rangle$ and $\left| \text{B} \right\rangle$ are orthogonal \cite{Resch}.

This points to the conclusion that some photons were shifted at the detector in one direction while \emph{others} in the opposite direction, as there are only three possible, mutually exclusive, stories: (1) each photon has simultaneously been near A and B, which was ruled out, (2) no photons have been near A and no photons have been near B, which is our position, (3) some photons have been near A and other photons have been near B.

But standard quantum mechanics tells us a very different story. The initial peaks corresponding to the different vibrational frequencies of A and B were a result of a non-zero probability amplitude at F. This is now zero since complete destructive interference has been restored. The peaks therefore disappear. From the Schrodinger evolution, there is no shift for some photons at the detector, due to mirror A, offsetting an opposite one for other photons due to mirror B. And ultimately, nothing to suggest that any photons have been near mirror A or B.

We now turn to FIG. \ref{fig: figure} (b), where the lower path is blocked. According to the TSVF, both the forward-evolving and the backward-evolving states are non-zero at mirror A and at mirror B, which vibrate at different frequencies---yet no peaks are observed! The authors attribute this to an insufficient number of photons reaching the detector. But even if the detector were sensitive enough, the TSVF cannot explain their size simply.

In the standard framework of quantum mechanics, when the lower path is blocked, the intensity of the light reaching the detector is proportional to the square of (the modulus of) the probability amplitude leaking from the inner interferometer (reflected from F), which is very small. But when the lower path is not blocked, interference with the state reflected from C results in an intensity at the detector that has a term linear in (the modulus of) the probability amplitude leaking from the inner interferometer, not as small, which is how we got the peaks in the first place.

In summary, if weak measurements are performed such that complete destructive interference is maintained, claims such as ``some photons have been inside the inner interferometer but they never entered and never left'' should not arise. Here, standard quantum mechanics not only explains the physics better than the two-state vector formulation the authors advocate, but it can tell a very different story.

Hatim Salih

Qubet Research, London NW6 1RE, UK

\begin{acknowledgments}
The author thanks Onur Hosten and Brian D. Josephson for useful comments. This work is partially supported by Qubet Research, a start-up in quantum information.
\end{acknowledgments}

\clearpage


\begin{thebibliography}{99}
\bibitem{Contrary} Contrary to the authors' claim in \cite{Reply}, mirrors A and B have to be rotated out of phase in order to restore complete overlap of the beams, therefore restoring complete destructive interference.
\end{thebibliography}
\end{document}